\documentclass[conference]{IEEEtran}
\IEEEoverridecommandlockouts
\usepackage{cite}
\usepackage[colorlinks,linkcolor=red,anchorcolor=blue,citecolor=blue]{hyperref}
\usepackage{amsmath,amssymb,amsfonts}
\usepackage{algorithmic}
\usepackage{graphicx}
\usepackage{subcaption}
\usepackage{textcomp}
\usepackage{booktabs}
\usepackage{threeparttable}
\usepackage{xcolor}
\usepackage{amsmath}
\usepackage{CJK}
\usepackage{svg}
\usepackage{caption}
\usepackage{multirow}
\usepackage{multicol}
\usepackage{mathtools}

\begin{document}

\title{
CDM-QTA: Quantized Training Acceleration for Efficient LoRA Fine-Tuning of Diffusion Model
\thanks{This work was supported in part by the Shenzhen Science and Technology Program 2023A007.}
}

\author{\IEEEauthorblockN{Jinming Lu}
\IEEEauthorblockA{\textit{Nanjing University}\\
Nanjing, China \\
jmlu@smail.nju.edu.cn}
\and
\IEEEauthorblockN{Minghao She}
\IEEEauthorblockA{\textit{Nanjing University}\\
Nanjing, China \\
mhshe@smail.nju.edu.cn}
\and
\IEEEauthorblockN{Wendong Mao}
\IEEEauthorblockA{\textit{Sun Yat-Sen University} \\
Shenzhen, China \\
maowd@mail.sysu.edu.cn}
\and
\IEEEauthorblockN{Zhongfeng Wang}
\IEEEauthorblockA{\textit{Sun Yat-Sen University} \\
Shenzhen, China \\
wangzf83@mail.sysu.edu.cn}
\thanks{Coreesponding author: Wendong Mao, Zhongfeng Wang}
}

\maketitle
\begin{abstract}
   Fine-tuning large diffusion models for custom applications demands substantial power and time, which poses significant challenges for efficient implementation on mobile devices. In this paper, we develop a novel training accelerator specifically for Low-Rank Adaptation (LoRA) of diffusion models, aiming to streamline the process and reduce computational complexity. By leveraging a fully quantized training scheme for LoRA fine-tuning, we achieve substantial reductions in memory usage and power consumption while maintaining high model fidelity. The proposed accelerator features flexible dataflow, enabling high utilization for irregular and variable tensor shapes during the LoRA process. Experimental results show up to 1.81\texttimes\  training speedup and 5.50\texttimes\  energy efficiency improvements compared to the baseline, with minimal impact on image generation quality.
    \end{abstract}


    \IEEEpeerreviewmaketitle

    \section{Introduction}
    Diffusion models have achieved remarkable success in image generation and artistic creation, allowing users to generate high-quality images from simple text prompts. These systems are capable of generating a vast array of objects, styles, and scenes—almost ``anything and everything"\cite{ramesh2022hierarchical,rombach2022high,saharia2022photorealistic,yu2022scaling}. As a versatile class of generative models, diffusion models have demonstrated notable  capabilities across a variety of applications, including image super-resolution\cite{saharia2022image,li2022srdiff}, inpainting\cite{song2020score}, shape generation\cite{cai2020learning}, image-to-image translation\cite{sasaki2021unit}, and molecular conformation generation\cite{xu2022geodiff}.

    However, despite their broad and general capabilities, users often wish to synthesize specific concepts based on their personal experiences, such as family members, pets or personal items. These concepts are not encountered during the large-scale pre-training procedure. Describing such concepts through text can be cumbersome, and most generative models struggle to reproduce these personal concepts with sufficient fidelity, which has increased demand for model customization \cite{kumari2023multi}.

    \begin{figure}
        \includegraphics[width=0.5\textwidth]{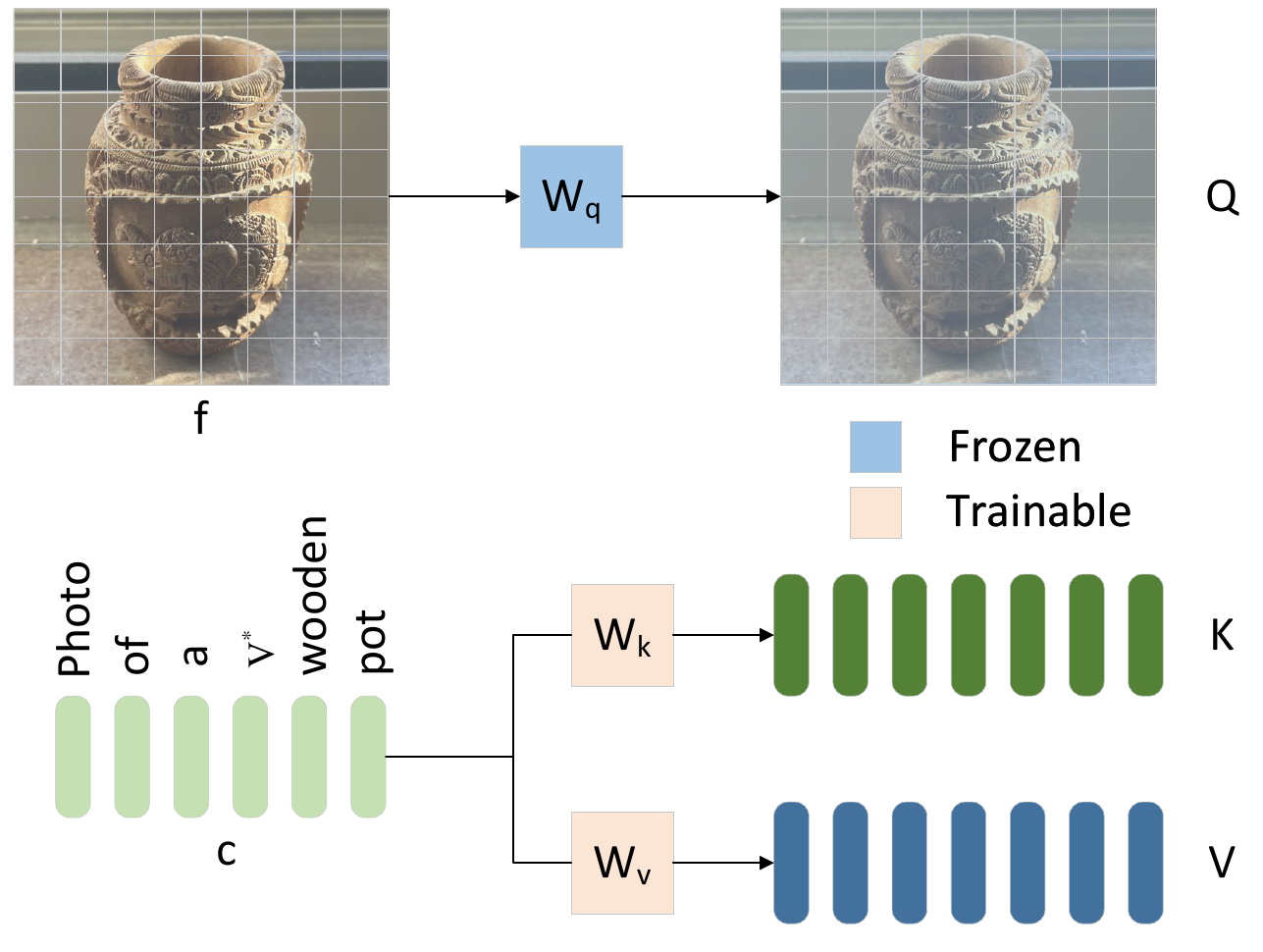}
        \caption{Cross-Attention module in the custom diffusion model.}
        \label{fig:cust}
    \end{figure}

    Custom Diffusion \cite{kumari2023multi} was proposed to enhance existing text-to-image diffusion models using a few user-provided images to incorporate new concepts. The fine-tuned model is then capable of generating new variations with existing concepts. Specifically, a small subset of model weights is identified, namely the key and value mappings from text to latent features in the cross-attention layers, while the rest remain frozen and do not participate in updates. To prevent model forgetting, a small set of real images with similar captions is used as target images.

    In traditional deep neural networks (DNNs) training, 32-bit single precision floating-point (FP32) has been the default across many DNN training frameworks and hardware systems. Although only 5\% of the weight are updated during fine-tuning, the frozen weights still participate in computation in subsequent steps. Therefore, despite the power of diffusion models, their application is limited by the massive number of parameters and computational complexity. For example, running Stable Diffusion\cite{rombach2022high} requires 16GB of memory and GPUs with over 10GB of VRAM, which is impractical for most consumer-grade PCs, let alone resource-constrained edge devices.

    To address the above challenges, we proposed an efficient hardware accelerator for custom diffusion model.  Our contributions are summarized as
    follows.
    \begin{enumerate}
        \item We propose an efficient fine-tuning method based on Low rank adaptation (LoRA) \cite{hu2021lora}, designed to expedite the concept fusion process. Subsequently, a quantized training method is developed to reduce computational resources and memory demands significantly, facilitating the implementation of integer calculations during training.
        \item We design a flexible hardware accelerator featuring a configurable dataflow that supports both weight stationary (WS) and output stationary (OS) modes. This flexibility allows efficient processing of irregular and small tensor computations in LoRA custom diffusion.
        \item Our experimental evaluation shows up to 1.81\texttimes\ training speedup and 5.50\texttimes\ energy efficiency improvement over the baseline architecture. Our design achieves 1.64\texttimes\  and 1.83\texttimes\ and improvements in terms of energy efficiency and area efficiency, respectively, compared to previous work.
    \end{enumerate}

    \begin{figure}[tp]
        \centering
        \includegraphics[width=0.8\linewidth]{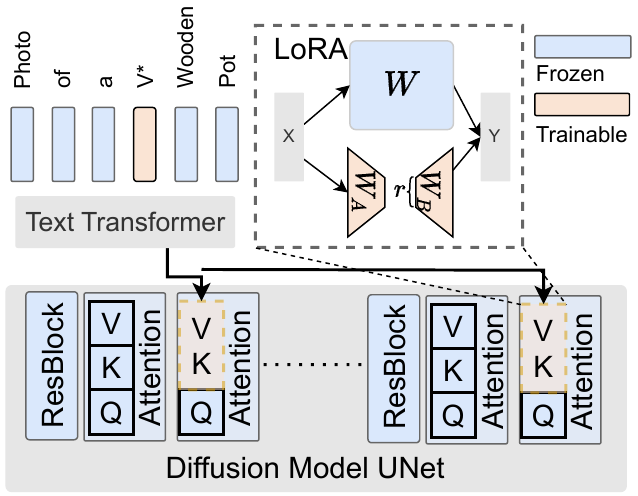}
        \caption{LoRA fine-tuning for Custom Diffusion model. Only weights in {\colorbox[HTML]{FAE4D4} {\textbf{pink}}} color are trainable, which accounts for a accounts for a tiny fraction of the entire model.}
        \label{fig:lora}
    \end{figure}

    \section{Algorithm}

    In this section, we introduce our comprehensive compression scheme designed to optimize the performance and efficiency of diffusion models. Our approach  consists of  two key components: a fine-tuning scheme leveraging Low-Rank Adaptation (LoRA) and a fully quantized scheme. These components are engineered to reduce computational demands while maintaining or improving model accuracy and output quality. This dual strategy streamlines the fine-tuning process, making it feasible  to deploy diffusion model in resource-constrained environments.

    \subsection{Fine-Tuning Scheme Based on LoRA}
    As shown in Figure \ref{fig:cust}, during the fine-tuning process of Custom Diffusion models, a new modifier token, \textbf{$V^*$} is introduced in front of the category name.  This fine-tuning primarily optimizes the key and value projection matrices within the cross-attention layers of the diffusion model, alongside the modifier token. The layers involved in optimization are referred to  as \textbf{non-frozen layers}, while those that do not participate are termed \textbf{frozen layers}. Consequently, the challenge of compressing the concept fusion process is translated into optimizing these layers using fewer resources.

    \begin{equation}
        \mathbf{Y} = \mathbf{XW} + \mathbf{XAB}^{T},
        \label{lora_formula}
    \end{equation}
    To address this, we have implemented a substitution of some non-frozen layers with Low-Rank Adaptation (LoRA). As shown in Figure \ref{fig:lora}, this adaptation redefines the original linear transformation in the cross-attention layers as Eq. (\ref{lora_formula}), where $\textbf{X} \in \mathbb{R}^{n \times d_{1}}, \textbf{W} \in \mathbb{R}^{d_{1} \times d_{2}}, \textbf{A} \in \mathbb{R}^{d_{1} \times r}, \textbf{B} \in \mathbb{R}^{d_{2} \times r}$, and $\textbf{r} \ll min({d_{1},d_{2}})$. Accordingly, the update of the large size weight matrix \textbf{W} is converted to the update of two low-rank matrices \textbf{A} and \textbf{B}.
    As a result, only 5\% of total parameters actively participate in updates, leading to significant savings in computational resources and memory costs. Moreover, given the inherent support for LoRA within the diffusion model framework, replacing parts of the model with LoRA does not result in a substantial loss of accuracy.

    \begin{figure}[tp]
        \centering
        \includegraphics[width=0.8\linewidth]{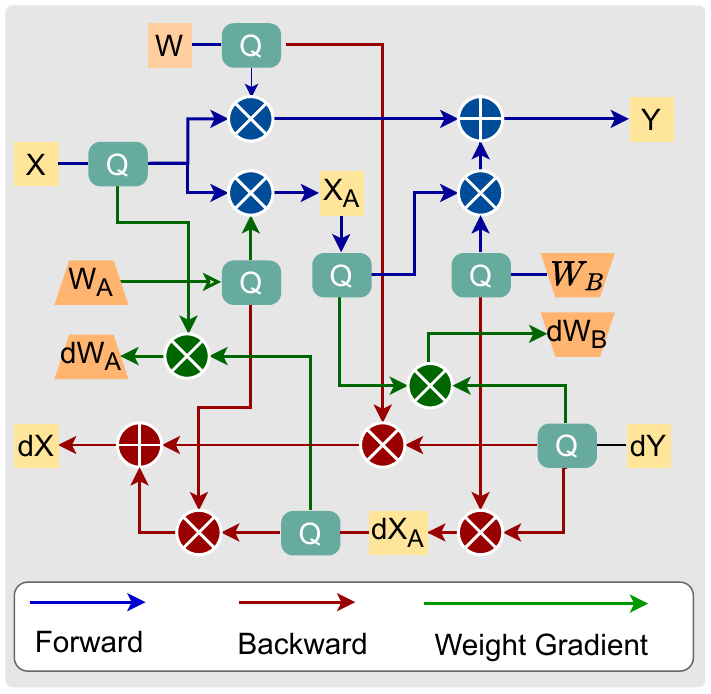}
        \caption{Mixed precision quantization scheme based on LoRA}
        \label{fig:quantscheme}
    \end{figure}

    \begin{figure*}[tp]
        \centering
        \begin{subfigure}{0.3\textwidth}
            \centering
            \includegraphics[width=\textwidth]{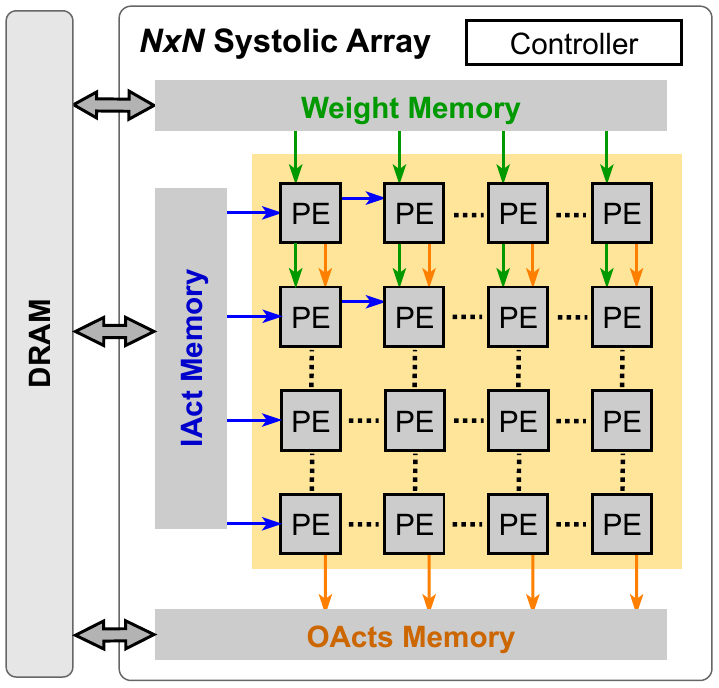}
            \caption{System architecture.}
            \label{fig:sys_arch}
        \end{subfigure}
        \begin{subfigure}{0.25\textwidth}
            \centering
            \includegraphics[width=\textwidth]{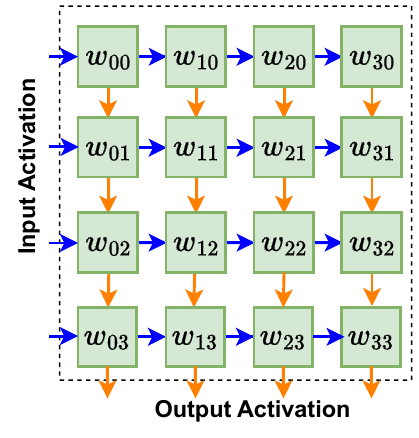}
            \caption{WS dataflow.}
            \label{fig:ws}
        \end{subfigure}
        \begin{subfigure}{0.25\textwidth}
            \centering
            \includegraphics[width=\textwidth]{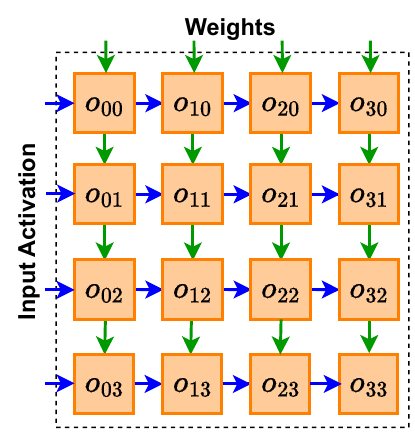}
            \caption{OS dataflow.}
            \label{fig:os}
        \end{subfigure}
        \caption{Overview of hardware architecture and dataflow.(a) The hardware architecture of the proposed accelerator. (b) and (c) are the WS and OS dataflows for various computation processes.}
    \end{figure*}

    \subsection{Fully Quantized Training Scheme}
    Even though the training parameters are significantly reduced after applying customization LoRA fine-tuning for diffusion model, the overall amount of MAC operations and memory consumption of weights and intermediate data are still relatively high.
    The overall computing graph during training is shown in Fig. \ref{fig:quantscheme}, from which we can find that frozen weight still participate in the computation of backward propagation.

    To further reduce the complexity, we introduce a fully quantized training approach, where weights, activation, and gradients are all quantized into 8-bit integer format (INT8).
    To ensure the training convergence, a per-tensor quantization scheme is applied to weights, and per-channel/per-column  quantization scheme is applied to activations and gradients. The quantization process is written as Eq. \ref{eq:quant}.

    \begin{equation}
        \label{eq:quant}
        \begin{aligned}
            &S = \frac{X_{\text{max}}}{2^{q-1} - 1}, \\
            &\widetilde{X}  = \text{round}\left( \frac{X}{S} \right)\times S. \\
        \end{aligned}
    \end{equation}
\vspace{-0.3in}

    \section{ Proposed Flexible Hardware Design}

    \subsection{Hardware Architecture}
    The system architecture is presented in Fig. \ref{fig:sys_arch}, which consists of a control module, an $N\times N$ systolic array-based compute module, and SRAM memories. The control module is responsible for receiving  instructions and configurations,  while coordinating the operations of other modules.
    The compute module is configurable to perform General Matrix Multiplications (GEMM) using both WS and OS dataflows.
    The SRAM memories store weight, input, and output tensors, which are fetched from off-chip DRAM.

    In our implementation, the systolic array is sized at  $64\times 64$. The memory capacities are 512KB, each for input and weight memory, 1MB for output memory.
    To reduce the latency caused by external memory access, a double buffer technique is employed.

    \subsection{Dataflow}
    {
        In the custom diffusion model, the cross-attention layers combine text prompt embeddings with image features. However, the sequence length of text embedding is small ($<77$), while the sequence length of images is large (4096), which causes significant variations in computational characteristics. The introduction of low-rank weights from LoRA exacerbates this issue. If not handled properly, hardware computing efficiency will be significantly compromised. }
    Therefore, our accelerator is designed to support both WS and OS dataflows on a unified PE array. The different dataflow modes are illustrated in Fig. \ref{fig:ws} and \ref{fig:os}.

    \textbf{WS}: Using WS dataflow, GEMMs are executed in an inner product manner. Weight vectors are first loaded into the PE array and stored locally in registers of each PE  for reuse. Input vectors are then streamed into the PE rows from left to right, and propagate in a systolic fashion. Outputs are collected from the bottom PE array row and aligned to form output vectors.

    \textbf{OS}: The OS dataflow performs GEMMs in an outer product fashion. A pair of input and weight vectors are fetched to generate $N\times N$ output partial sums. The input and weight vectors are broadcast across the PE array horizontally and vertically, respectively. The partial sums are accumulated temporally in the PEs and streamed out once accumulation completes. These outputs are then stored in output memory.

    The WS and OS dataflows employ different schemes for data propagation and partial sum accumulation, which leads to variations in PE utilization and memory traffic. By selecting the optimal dataflow for each layer, overall performance can be improved significantly.

    \section{Experiments}

    \subsection{Evaluation Methodology}
    In this section, we present the results of our method across  multiple datasets using the  Stable Diffusion model.  We show both qualitative results, demonstrating the effects of our solution on generating images, and quantitative results, comparing power consumption and computing resource usage.

    \subsubsection{Algorithm }
    Following the experimental design in Custom Diffusion, we conducted experiments on multiple target datasets spanning various categories, including scenes, pets, and objects.
    \subsubsection{Hardware}
    We implement the accelerator in System Verilog RTL. The RTL design was synthesized using Synopsys Design Compiler with 45nm FreePDK technology\cite{freepdk45}  to obtain the area and power consumption. We use CACTI 7.0 \cite{cacti} to model the energy and area consumption of SRAM buffers.
    A cycle-level simulator was developed  based on SCALE-Sim \cite{scalesim} to determine the optimal dataflow configurations.

    \subsection{Qualitative evaluation}

    \begin{figure}[htbp]
        \centering
        \begin{subfigure}[t]{0.15\textwidth}
            \centering
            \includegraphics[width=\textwidth]{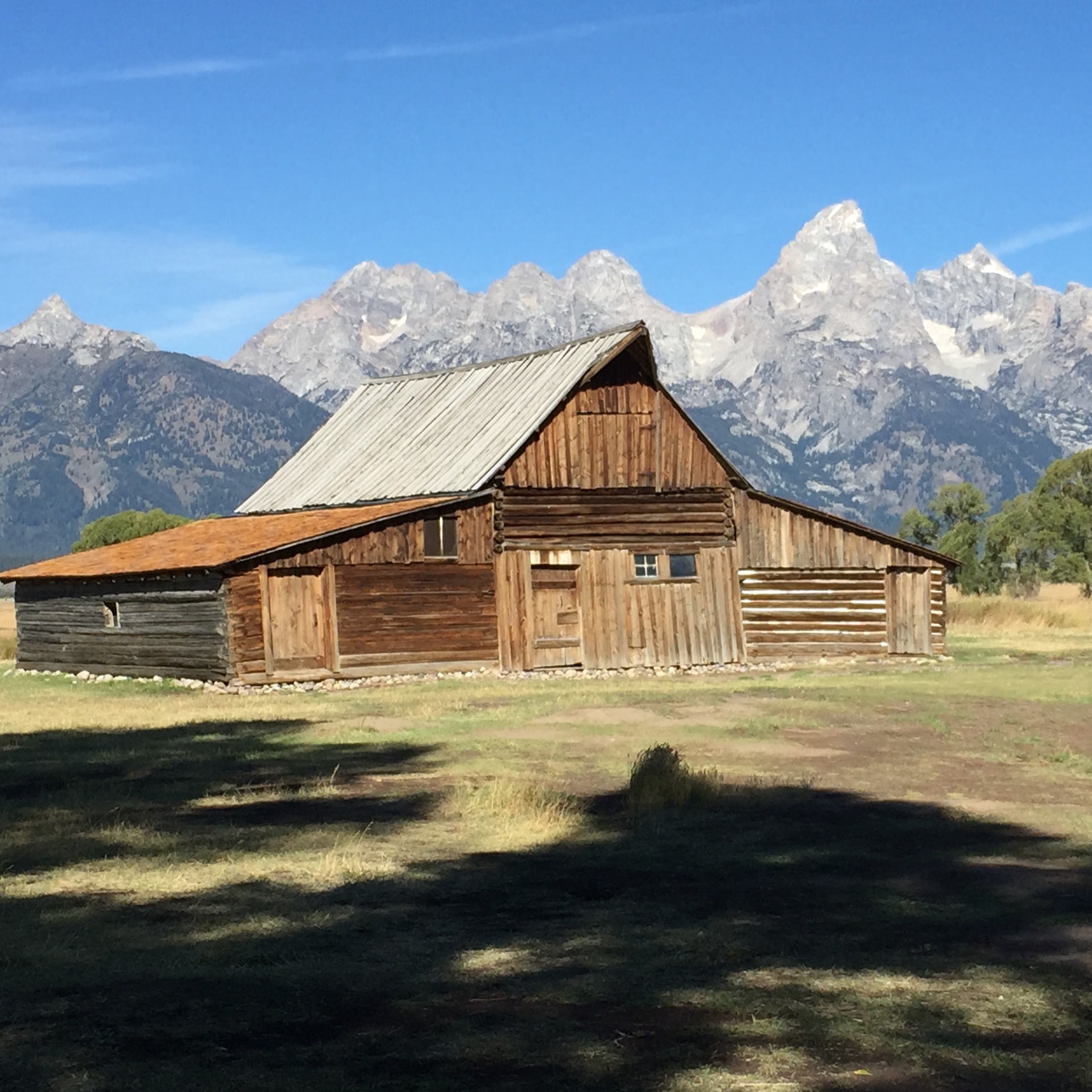}
            \caption{barn}
        \end{subfigure}
        \hfill
        \begin{subfigure}[t]{0.15\textwidth}
            \centering
            \includegraphics[width=\textwidth]{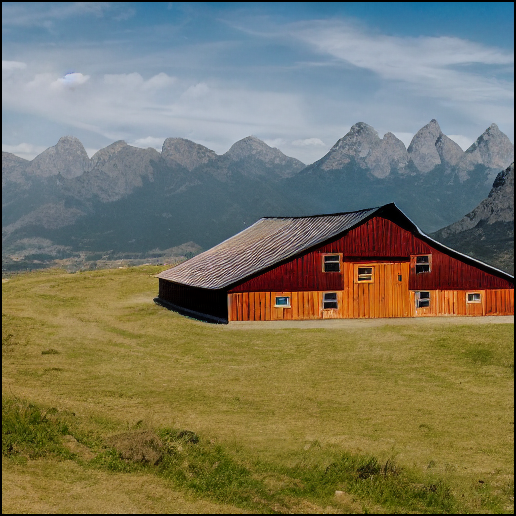}
            \caption{$\langle new\rangle$ barn near the mountain}
        \end{subfigure}
        \hfill
        \begin{subfigure}[t]{0.15\textwidth}
            \centering
            \includegraphics[width=\textwidth]{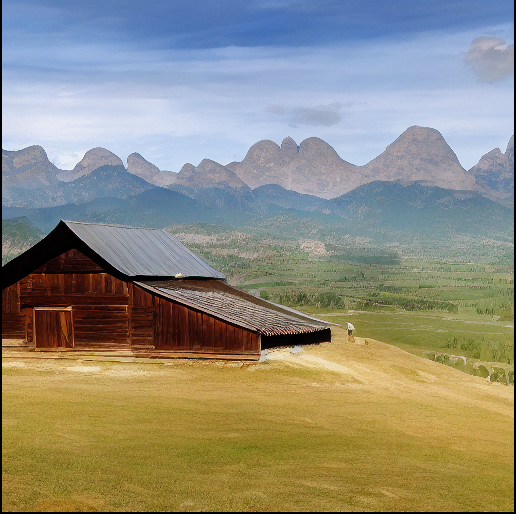}
            \caption{ours: $\langle new\rangle$ barn near the moutain}
        \end{subfigure}

        \vfill
        \begin{subfigure}[t]{0.15\textwidth}
            \centering
            \includegraphics[width=\textwidth]{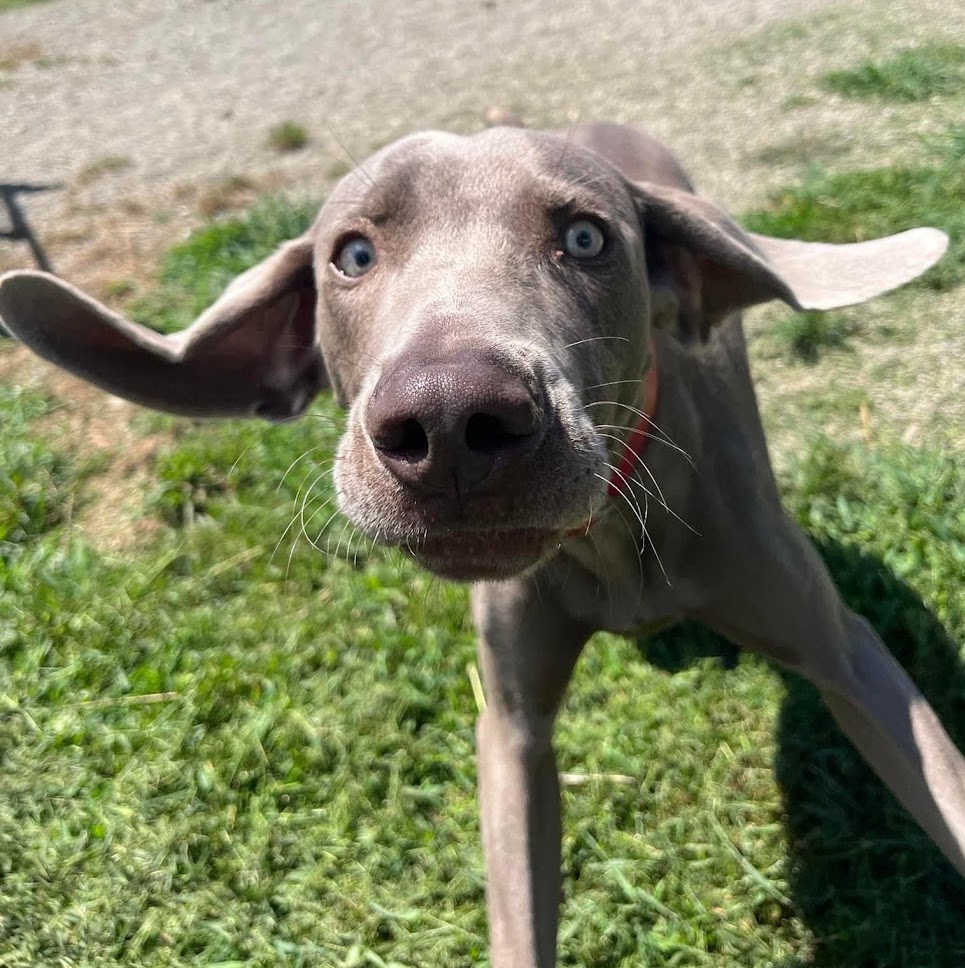}
            \caption{dog}
        \end{subfigure}
        \hfill
        \begin{subfigure}[t]{0.15\textwidth}
            \centering
            \includegraphics[width=\textwidth]{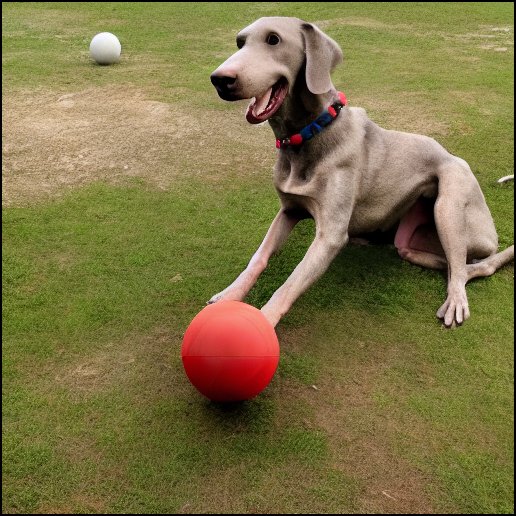}
            \caption{$\langle new\rangle$ dog playing a ball}
        \end{subfigure}
        \hfill
        \begin{subfigure}[t]{0.15\textwidth}
            \centering
            \includegraphics[width=\textwidth]{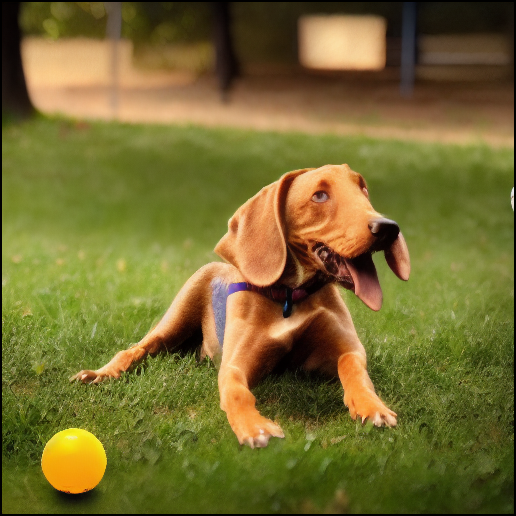}
            \caption{ours: a yellow $\langle new\rangle$ dog playing a ball}
        \end{subfigure}
        \vfill
        \begin{subfigure}[t]{0.15\textwidth}
            \centering
            \includegraphics[width=\textwidth]{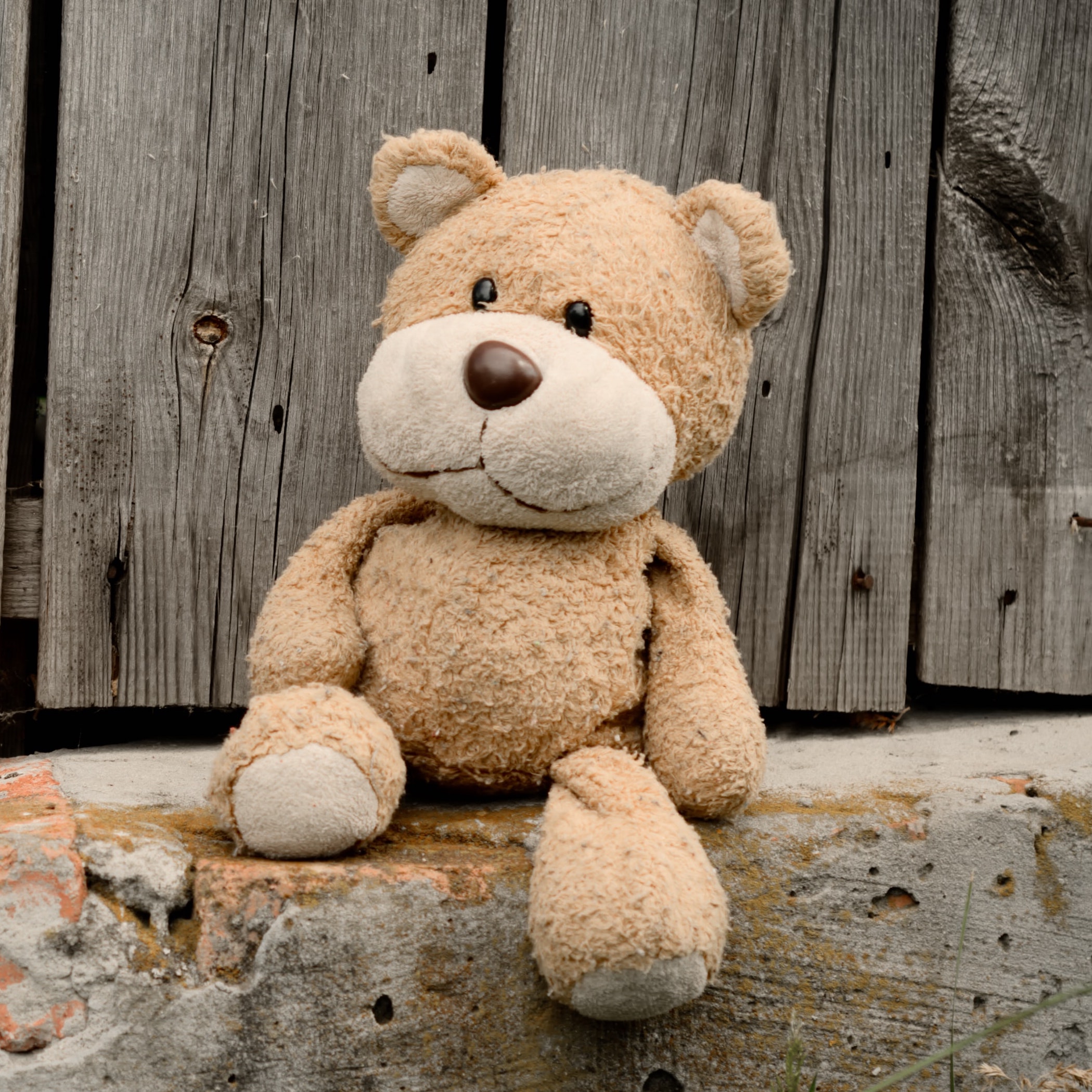}
            \caption{teddy-bear}
        \end{subfigure}
        \hfill
        \begin{subfigure}[t]{0.15\textwidth}
            \centering
            \includegraphics[width=\textwidth]{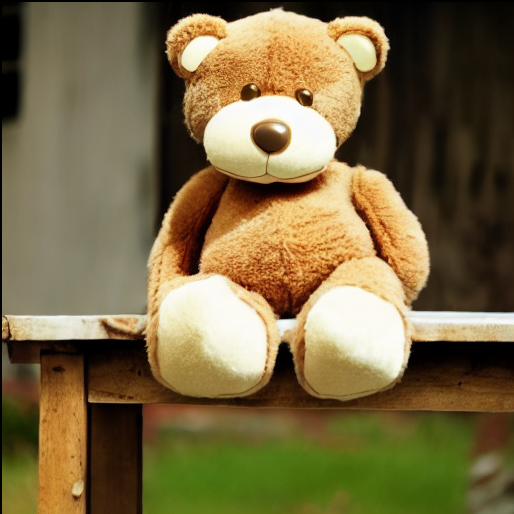}
            \caption{$\langle new\rangle$ teddybear sitting on the bench}
        \end{subfigure}
        \hfill
        \begin{subfigure}[t]{0.15\textwidth}
            \centering
            \includegraphics[width=\textwidth]{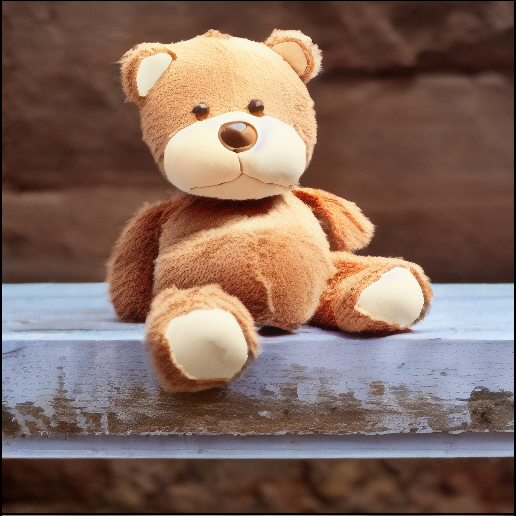}
            \caption{ours: $\langle new\rangle$ teddybear with smooth fur sitting on the bench}
        \end{subfigure}
        \caption{Comparison of the generation effects of custom diffusion and the quantitative compression model in this article}
        \label{fig:grid}
    \end{figure}

    In Fig. \ref{fig:grid}, we compare the image generation effects of the original Custom Diffusion method and our quantized compression model. We test each fine-tuned model using a set of prompts to evaluate the integration of target concepts into new scenes and the modification of target concept properties, such as e.g., color, shape.  Column 2 and 3 of Fig.  \ref{fig:grid} present  sample generations from both Custom Diffusion and our method. Our method demonstrates similar text-to-image alignment, captures visual details of the target object, and effectively fuses concepts, all while maintaining lower model storage requirements.

    \subsection{Hardware Performance}
    \begin{figure}[tp]
        \centering
        \begin{subfigure}{0.24\textwidth}
            \centering
            \includegraphics[width=\textwidth]{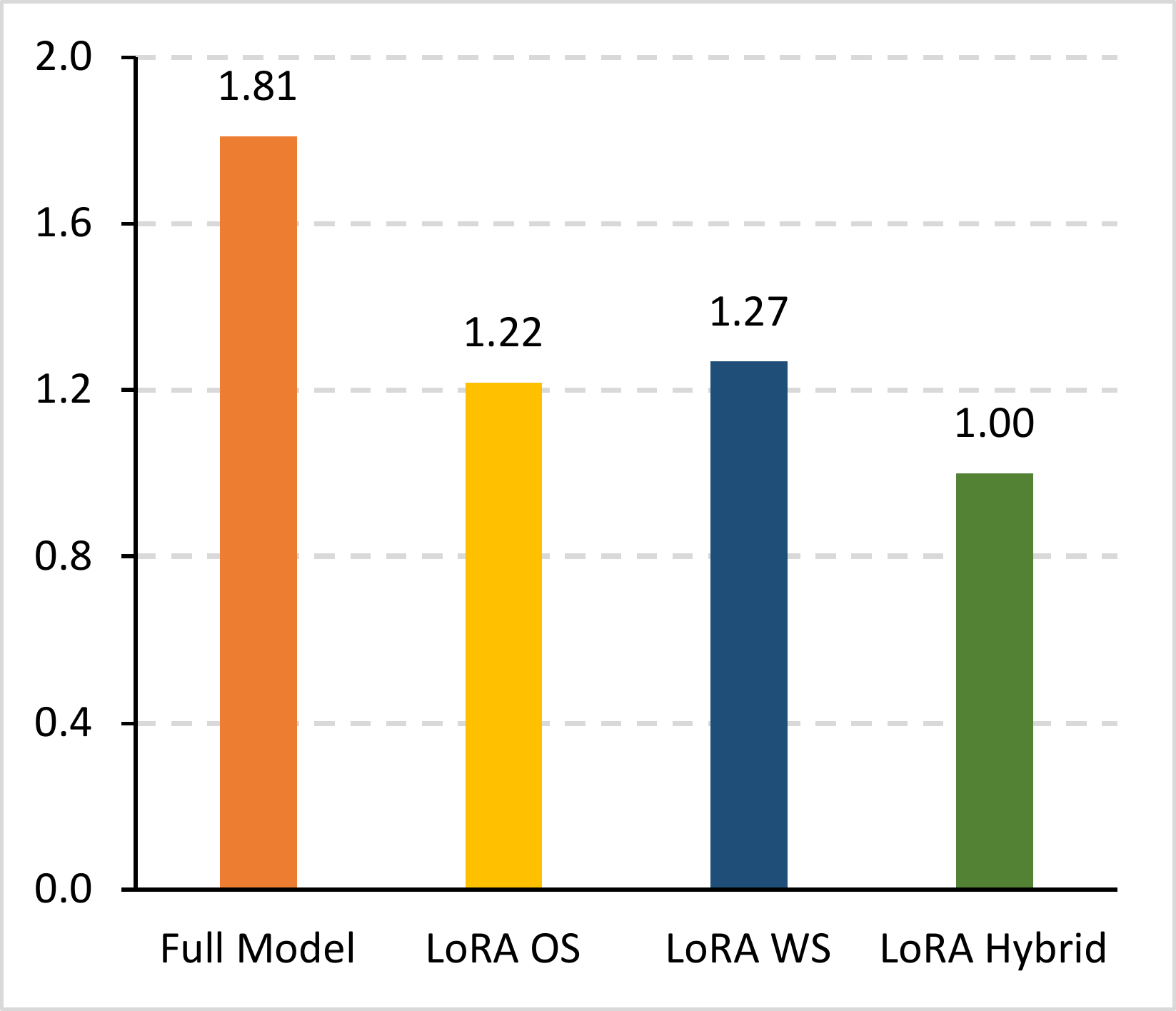}
            \caption{Latency}
            \label{fig:latency}
        \end{subfigure}
        \begin{subfigure}{0.24\textwidth}
            \centering
            \includegraphics[width=\textwidth]{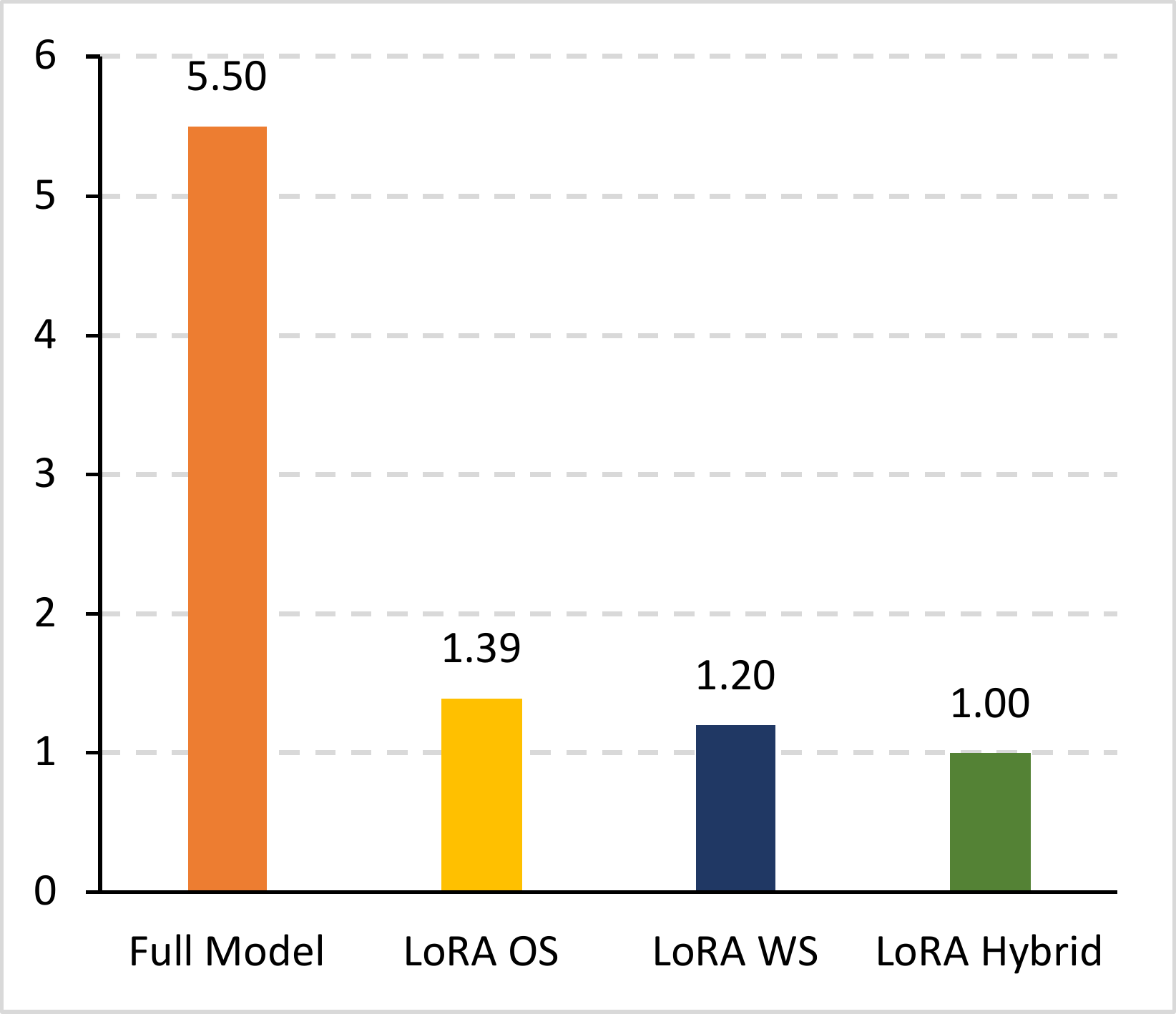}
            \caption{Energy Delay Product}
            \label{fig:edp}
        \end{subfigure}
        \caption{Performance and energy comparison.}
        \label{fig:comp}
    \end{figure}

    \begin{table}[]
    \centering
    \caption{Comparison with Previous Work}
    \label{tab:my-table}
    \begin{threeparttable}
    \resizebox{0.9\columnwidth}{!}{%
    \begin{tabular}{lcc}
    \toprule
    Work                          & \cite{liu2024fully}                                    & ours  \\ \midrule
    Technology (nm)               & 28                                                     & 45    \\
    Voltage (V)                   & 0.9                                                    & 1.1   \\
    Frequency (MHz)               & 400                                                    & 400   \\
    Area (mm$^2$)                 & 1.89                                                   & 15.07 \\
    Power (W)                     & 0.61                                                   & 3.49  \\
    Performance (TOPS)            & 0.55                                                   & 3.28  \\
    Energy Efficiency (TOPS/W)    & 0.90 (0.57*)                                            & 0.94  \\
    Area Efficiency (TOPS/mm$^2$) & 0.29 (0.12*)                                            & 0.22  \\ \bottomrule
    \end{tabular}
    }
    \begin{tablenotes}
        \footnotesize
        \item[*] Scaled to 45nm.
    \end{tablenotes}
    \end{threeparttable}%
    \end{table}

    TABLE \ref{tab:my-table} compares the proposed accelerator with previous designs for diffusion models. The proposed design achieves a peak performance of 3.28 TOPS (Tera Operations Per Second) while consuming 3.49W of power. The design occupies an area of 15.07mm², translating into an area efficiency of 0.22 TOPS/mm². It achieves a good balance between computational performance and power usage, showing 1.64\texttimes\  and 1.83\texttimes\ and improvements in terms of energy efficiency and area efficiency compared with \cite{liu2024fully}.

    Fig. \ref{fig:comp} illustrates performance comparisons between different configurations: Full Model, LoRA OS, LoRA WS, and LoRA Hybrid. Two key metrics are evaluated: latency and energy delay product (EDP).

    As shown in Fig. \ref{fig:latency}, our LoRA Hybrid configuration provides a 1.81\texttimes\ speedup over the full model baseline.
    When compared to LoRA OS and LoRA WS, which use fixed dataflows, the hybrid dataflow achieves speedups of 1.22\texttimes\ and 1.27\texttimes, respectively.
    Fig. \ref{fig:edp} demonstrates that LoRA Hybrid design achieves an EDP reduction of 5.5\texttimes, 1.39\texttimes, and 1.20\texttimes\  over the full model, LoRA OS, and LoRA WS, respectively.
    These results indicates the superiority of our design  in both performance and energy efficiency.

\section{Conclusion}
Based on the LoRA fine-tuning scheme and the proposed fully quantized method, we optimized  custom diffusion models to significantly reduce computing resource requirements  and memory consumption.
The combination of these optimization schemes enables diffusion models to achieve higher efficiency and performance in both the training and inference phases.
Moreover, we validated the effectiveness of our algorithms on hardware platforms, demonstrating  that our optimizations not only perform well in theoretical simulations but also translate into tangible benefits in real-world applications. Hardware evaluations demonstrates that our approach can reliably achieves up to 1.81\texttimes\ processing speed  and 5.4\texttimes\ improvement in energy efficiency , paving the way for broader deployment in practical scenarios.

\newpage
\bibliographystyle{ieeetr}
\bibliography{ref}

\end{document}